\documentclass[11pt]{article}
\usepackage{amsmath,amssymb,epsfig}
\title{Oversampled Adaptive Sensing}

\newcommand{\be}{\begin{equation}}
\newcommand{\ee}{\end{equation}}
\newcommand{\bea}{\begin{eqnarray}}
\newcommand{\eea}{\end{eqnarray}}
\def\expect{\mathop{\mbox{$\mathsf{E}$}}}
\def\mathlette#1#2{{\mathchoice{\mbox{#1$\displaystyle #2$}}
                               {\mbox{#1$\textstyle #2$}}
                               {\mbox{#1$\scriptstyle #2$}}
                               {\mbox{#1$\scriptscriptstyle #2$}}}}
\def\argmax{\mathop{\rm argmax}}

\newcommand{\matr}[1]{\mathlette{\boldmath}{#1}}
\newcommand{\matt}[1]{\mathlette{\boldmath}{\tilde #1}}
\newcommand{\R}{\mathbb R}

\newcommand{\Tb}{T_{\rm b}}
\newcommand{\Tm}{T_{\rm m}}
\newcommand{\xnk}{x_{n_m}}
\newcommand{\xnkq}{x_{n(m)}^2}
\newcommand{\kk}{{\cal M}_m}
\newcommand{\myk}{\matt y_m}
\newcommand{\mykp}{\matt y_{m+1}}
\newcommand{\mse}{{\rm MSE}}

\usepackage[T1]{fontenc}
\usepackage[utf8]{inputenc}
\usepackage{authblk}

\author[1]{Ralf R.\ M\"uller}
\author[1]{Ali Bereyhi}
\author[2]{Christoph F.\ Mecklenbr\"auker}
\affil[1]{\small Institute for Digital Communications\\
FAU Erlangen-N\"urnberg, Germany}
\affil[2]{Institute of Telecommunications\\
Technische Universit\"at Wien, Austria}

\date{}

\begin{document}

\maketitle

\begin{abstract}
We develop a Bayesian framework for sensing which adapts the sensing time and/or basis functions to the instantaneous sensing quality measured in terms of the expected posterior mean-squared error. For sparse Gaussian sources a significant reduction in average sensing time and/or mean-squared error is achieved in comparison to non-adaptive sensing. For compression ratio 3, a sparse 10\% Gaussian source and equal average sensing times, the proposed method gains about 2 dB over the performance bound of optimum compressive sensing, about 3 dB over non-adaptive 3-fold oversampled orthogonal sensing and about 6 to 7 dB to LASSO-based recovery schemes while enjoying polynomial time complexity.

We utilize that in the presence of Gaussian noise the mean-squared error conditioned on the current observation is proportional to the derivative of the conditional mean estimate with respect to this observation.
\end{abstract}

\section{Introduction}

The term compressive sensing (CS) refers to methods to reduce the number of measurements below the number of components to be measured by utilizing the redundancy of the measured data. In the early days of CS, measurements were exclusively formed by linear combinations of the data and redundancy was restricted to sparsity of the data in some basis. While CS is now understood in a broader sense \cite{wu:10a}, linear measurements and sparsity still dominate the applications of CS. In the sequel, we will exclusively refer to linear measurements, i.e.\ the measurement vector is formed as
\begin{equation}
\label{om}
\matr y = \matr {Ax} + \matr z
\end{equation}
with $\matr x\in \R^{N\times 1}$, $\matr A\in\R^{K\times N}$ and the noise vector $\matr z\in\R^{K\times 1}$.

The goal is to estimate the data $\matr x$ given the measurements $\matr y$ subject to some non-negative distortion measure
\be
{d} [\matr x,r(\matr y)]
\ee
where the reconstruction function $r(\matr y)$ is to be chosen such that the
conditional average distortion
\be
D(\matr y) = \expect\limits_{\matr x|\matr y} {d} [\matr x,r(\matr y)]
\ee
is minimized.

In order to achieve good performance in terms of reconstruction fidelity and compression rate, the coefficients of the linear measurements should fulfill the restricted isometry property (RIP) \cite{candes:05}. The RIP ensures that all data components actually enter the measurements in a similar way. In practice, this is often implemented by a pseudorandom choice of the measurement matrix $\matr A$.

Measurements can be performed in parallel or sequentially. For parallel measurements, $K$ analog-to-digital converters (ADCs) are required, while for sequential measurements a single ADC can do. For sequential measurements, the current linear combination may depend on the outcome of previous measurements.
This allows to compromise on the RIP. If previous measurements allow to infer that certain components of the data are already known to sufficient accuracy, future measurements, i.e.\ linear combinations need not include them.
Moreover, sequential measurements allow to adapt the time spent on an individual measurement $y_k$ to depend on the previous measurements $[y_1,\dots,y_{k-1}]$.

Hybrid forms of parallel and sequential measurements are also possible.
In that case, several measurements are performed in parallel. Then, the measurement matrix is adapted and some other parallel measurements follow. We will see in the sequel, that such a hybrid procedure is most advantageous for many applications.
 
Adaptation of measurement time and adaptation of linear combinations can be put under a single more general umbrella: Let $\Tb$ denote the time required for all the measurements of data vector $\matr x$. Then, $\Tm=\Tb/K$ is the (average) time required per one out of $K$ sequential measurements. Let us, now, quantize the (adaptive) measurement time on a finer grid than $\Tm$ such that $T=\Tm/M$ for some integer $M>1$. Furthermore, let $t_k$ denote the duration of the $k^{\rm th}$ measurement.
Then, we can represent our $K$ measurements with adaptive timing and average duration $\Tm$ per sequential measurement, by $MK$ sequential measurements each with fixed duration $T$ by a modification of the measurement matrix $\matr A\in \R^{K\times N} \mapsto \matr {\tilde A}\in \R^{MK\times N}$ such that the $k^{\rm th}$ row of $\matr A$ is repeated $t_k/T$ times for all $k$ and we get the oversampled measurement vector
\be
\label{mm}
\matt y = \matt A\matr x + \matt z
\ee
with $\matt A\in\R^{MK\times N}$ and the noise vector $\matt z\in\R^{MK\times 1}$. Thus, time adaptation is a special case of oversampled adaptation of the measurement matrix.

Without time adaptation, i.e.\ $t_k = \Tm \forall k$ and independent white Gaussian noise of fixed power spectral density $N_0$, the measurements \eqref{mm} perform exactly identical to their counterparts in \eqref{om} in terms of measurement time and reconstruction error. The equivalence is obvious for the measurement time. Concerning the reconstruction error, one should note that \eqref{mm} can be obtained from \eqref{om} by means of repetition coding. It is well-known in coding theory that repetition coding does not gain over uncoded transmission in the presence of additive white Gaussian noise \cite{mackay:03}.

In contrast to non-adaptive CS, the modified measurement matrix $\matt A$ need not fulfill the RIP. In fact, as we will demonstrate later on, even a modified measurement matrix with a single nonzero element in each row, i.e.\ unit row weight performs well. 

Adaptation of the measurement matrix is a sophisticated task with a sparse literature list. 
The concept of Bayesian adaptive sensing was introduced in \cite{ji:08}. An iterative scheme to design the sequence of sensing vectors was proposed in \cite{haupt:09}, improved in \cite{haupt:09a} and \cite{malloy:14}, and applied to image compression in \cite{zhu:15}. Theoretical limits on adaptive compressive sensing were found in \cite{arias-castro:13}.

None of the above references allows for oversampling, however.
In the sequel, we will address this issue starting from the adaptation of the measurement time for measurement matrices with unit row weight.

%%%%%%%%%%%%%%%%%%%%%%%
\section{Sparsity Model}
Without loss of generality, we assume that the data vector $\matr x$ is sparse in the domain of the measurements, i.e.\ the sparsity level is given by the Hamming weight (zero-norm) of the data vector $\matr x$.
In practice, the data vector is often sparse in a different domain obtained by a linear transformation, e.g., the Fourier transform. In the latter case, the sparsity level would not be the Hamming weight of the data vector, but the Hamming weight of the data vector's Fourier transform.  Like any linear transform, the Fourier transform can be absorbed into the measurement matrix. Thus, sparsity in another domain is equivalent to a linear premultiplication of the measurement matrix. In the sequel, we assume that this linear premultiplication is already contained in the measurement matrix $\matr A$ in \eqref{om}.

%%%%%%%%%%%%%%%%%%%%%%%
\section{Unit Row Weight Measurements}

For measurement matrices with unit row weight, only a single component of the data vector $\matr x$ is sensed at a time. For sake of simplicity, let us consider the entries of the modified measurement matrix in \eqref{mm} to be chosen from the binary alphabet $\{0,1\}$.
Thus, we consider the channel 
\begin{equation}
\label{sgc}
\tilde y_m=\xnk+\tilde z_m
\ee
with the zero-mean white Gaussian noise $\tilde z_m$ of variance $\sigma^2=N_0/T = MN_0/\Tm$ and $n_m$ indicating which data symbol is measured at time instant $m$ out of a total of $MK$ oversampled measurements.
Let $\kk$ denote the set of all previous time instances $m^\prime$ that were used to measure $\xnk$, i.e.\
\be
\kk = \left\{ m^\prime \le m: n_{m^\prime} = n_m \right\}
\ee
and let the vector $\myk$ contain only those components of $\matt y$ whose indices are contained in $\kk$. 

With this notation, Bayes' law implies
\begin{align}
{\rm p}(\xnk|\myk) &= \frac{ {\rm p}(\myk| \xnk) {\rm p}(\xnk) } {{\rm p}(\myk)}\\
& = \frac{{\rm e}^{-\sum_{i\in\kk} (\tilde y_i - \xnk)^2/2\sigma^2}   {\rm p}(\xnk)}
{\int {\rm e}^{-\sum_{i\in\kk} (\tilde y_i - \xnk)^2/2\sigma^2}   {\rm dP}(\xnk)}.
\label{bayes1}
\end{align}

Let $r(\myk)$ denote the reconstruction function to estimate the data $\xnk$.
Thus, the mean square error $\mse(\myk)$ for a given observation $\myk$ reads
\begin{align}
\mse(\myk) &= \expect\limits_{\xnk|\myk}\left[\xnk - r(\myk)\right]^2\\
&= \int \left[\xnk - r(\myk)\right]^2 {\rm dP}(\xnk|\myk).
\label{mse}
\end{align}
Optimizing with respect to $r(\myk)$ gives
\begin{align}
0=\frac{\partial\mse(\myk)}{\partial r(\myk)} =2\int \left[r(\myk)-\xnk\right] {\rm dP}(\xnk|\myk)
\end{align}
implying
\be
\label{genrec}
r(\myk) = \int \xnk {\rm dP}(\xnk|\myk).
\ee
%%%%%%%%%%%%%
\subsection{Single Observations}
For single observations, a particularly helpful relation between the reconstruction function and the posterior MSE can be derived.
Consider the scalar Gaussian channel \eqref{sgc} for $|\kk|=1$ and drop the indices and tilde for sake of simplicity of notation. 

The derivative of the channel with respect to the observation is given by
\be
\frac{\partial {\rm p}(y|x)}{\partial y} = \frac{x-y}{\sigma^2} \, {\rm p}(y|x). 
\ee
With this result, the derivative of the likelihood function with respect to the observation is found to be
\begin{align}
\frac{\partial {\rm p}(x|y)}{\partial y} & = {\rm p}(x) \frac\partial{\partial y} \frac{{\rm p}(y|x)}{{\rm p}(y)}\\
&= {\rm p}(x) \frac{{\rm p}(y)\frac\partial{\partial y}{\rm p}(y|x) - {\rm p}(y|x) \frac\partial{\partial y}{\rm p}(y)}{{\rm p}^2(y)}\\
&=\frac{{\rm p}(x,y)}{\sigma^2} \frac{(x-y){\rm p}(y)-\int(x-y){\rm p}(y|x){\rm dP}(x) }{{\rm p}^2(y)}\\
&= \frac{{{\rm p}(x|y)}}{\sigma^2}
\left[ {
x - \int x {\rm dP}(x|y)
}\right].
\end{align}
Thus, we have
\begin{align}
\sigma^2 \frac{\partial r(y)}{\partial y} & = \sigma^2 \int\limits_{-\infty}^{\infty} x \frac{\partial {\rm p}(x|y)}{\partial y} {\rm d}x\\
&= \int x^2 {\rm dP}(x|y) - \left[\int x {\rm dP}(x|y)\right]^2\\
&= \mse(y).
\label{diffmse}
\end{align}
This implies
\begin{align}
\mse &= \int \mse(y) {\rm dP}(y) \\
&= \sigma^2 \int {\rm p}(y) {\rm d}r(y).
\end{align}
These relations will be helpful in the following subsections to derive mean-square distortions for various prior distributions.
%%%%%%%

\subsection{Gaussian sparse data}
Consider now the case that $\xnk$ follows a mixed distribution.
With probability $p$, it is 0. With probability $1-p$, it is Gaussian distributed with zero mean and unit variance for all $n_m$. Thus, each of the $N$ source symbols has average power $1-p$ and average energy $E_{\rm s} = (1-p)\Tm$.
Specializing \eqref{bayes1} to sparse Gaussian data, we get
\begin{align}
{\rm p}(\xnk|\myk) &= \frac{{\rm e}^{-\sum\limits_{i\in\kk}\frac{(\tilde y_i-\xnk)^2}{2\sigma^2}}{\rm p}(\xnk)}{\frac{1-p}{\sqrt{2\pi}} \int\limits_{-\infty}^{+\infty} {\rm e}^{-\sum\limits_i\frac{(\tilde y_i-\xi)^2}{2\sigma^2}} e^{-\frac{\xi^2}2}{\rm d}\xi + p e^{-\sum\limits_i\frac{\tilde y_i^2}{2\sigma^2}}}\\
&= \frac{{\rm e}^{-\sum\limits_{i\in\kk} \frac{\tilde y_i^2-2\tilde y_i \xnk+\xnkq}{2\sigma^2}}{\rm p}(\xnk)}{\frac{1-p}{\sqrt{1+\frac {|\kk|}{\sigma^2}} } {\rm e}^{-\sum\limits_i \tilde y_i^2 (|\kk|-1+\sigma^2) - \sum\limits_{j>i} \frac{ \tilde y_i\tilde y_j}{\sigma^2(|\kk|+\sigma^2)}}  + p {\rm e}^{-\sum\limits_i\frac{\tilde y_i^2}{2\sigma^2}}}\\
& =  \frac{{\rm e}^{\xnk\sum\limits_{i\in\kk} (2\tilde y_i-\xnk)/(2\sigma^2)}{\rm p}(\xnk)}{\frac{1-p}{\sqrt{1+\frac {|\kk|}{\sigma^2}}  }{\rm e}^{(\sum\limits_i  \tilde y_i)^2/(2\sigma^2(|\kk|+\sigma^2))}  + p}
\label{eq22}
\end{align}
where the sums, if not explicitly stated run over the set $\kk$.
This can be plugged into \eqref{genrec} yielding
\begin{align}
%\label{eq23}
r(\myk) &= \int \frac{\xnk{\rm e}^{\xnk\sum\limits_{i\in\kk} (2\tilde y_i-\xnk)/(2\sigma^2)}{\rm dP}(\xnk)}{\frac{1-p}{\sqrt{1+\frac {|\kk|}{\sigma^2}}  }{\rm e}^{(\sum\limits_i  \tilde y_i)^2/(2\sigma^2(|\kk|+\sigma^2))}  + p}\\
%\label{eq24}
& =  \frac{\frac {1}{|\kk|+\sigma^2} \sum_i\tilde y_i }
{1+\sqrt{1+\frac {|\kk|}{ \sigma^2}} \frac p{1-p} {\rm e}^{-\frac{(\sum_i\tilde y_i)^2}{2\sigma^2(|\kk|+\sigma^2)}}  }.
\label{recfun}
\end{align}
Note that the reconstruction function depends only on the average of the components of the vector $\myk$
\be
\label{defaverage}
\overline y_m = \frac1{|\kk|}\sum\limits_{i\in\kk} \tilde y_i.
\ee

Figure~\ref{figrecfun} 
\begin{figure}
\centerline{\epsfig{file=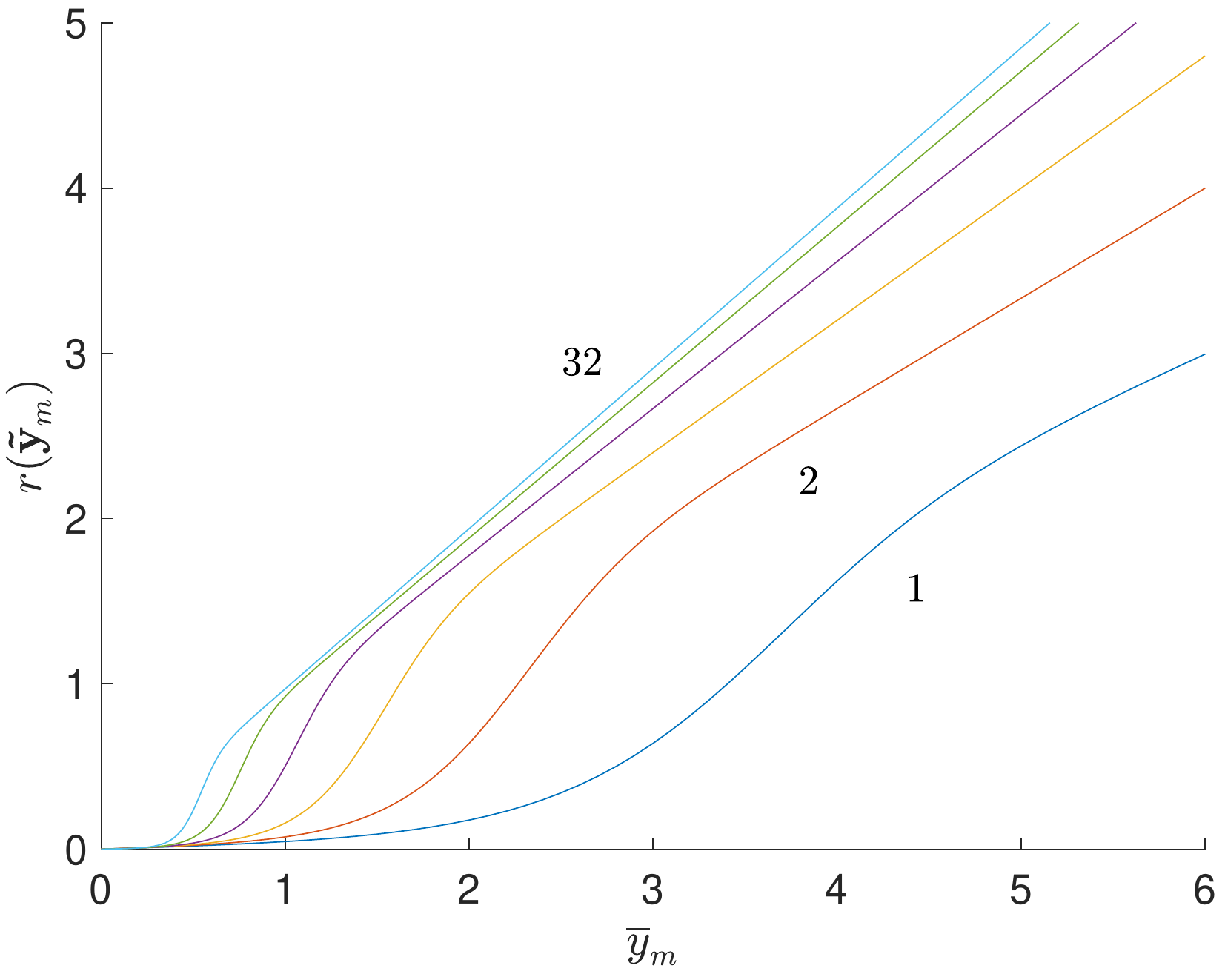,width=.75\columnwidth}}
\caption{\label{figrecfun}
Reconstruction function for sparse Gaussian source with $p=0.9$ and $\sigma^2=1$ for 1, 2, 3, 4, 8, 16, and 32 measurements shown by the lowest to highest curve, resp.}
\end{figure}
illustrates this reconstruction function. Only the positive part is shown as the function is an odd function, i.e.\ $r(-\myk) = - r(\myk)$.
For large arguments and large number of measurements, the reconstruction function approaches to the identity function. For only few, thus insecure, measurements, the reconstruction function attenuates the mean of the observations to combat the influence of the noise. For arguments close to zero, the reconstruction function shows a further attenuation effect, since such arguments induce a high probability for the sensed data to be zero. 

\newcommand{\e}{\tilde{e}}
\newcommand{\sq}{\sigma^2}
\newcommand{\df}{\tilde d}
We find the posterior MSE by means of \eqref{diffmse} as
\begin{align}
\mse(\myk) &= \frac{\sigma^2}{|\kk|} \frac{\partial r(\myk)}{\partial \overline y_m}\\
&= \frac{\sq + \frac{\df |\kk|^2\overline y_m^2}{(1+\df)(|\kk|+\sq)}}{(1+\df)(|\kk|+\sq)}
\label{msefun}
\end{align}
with the shortcuts
\begin{align}
\e & = e^{\frac{|\kk|^2\overline y_m^2}{2\sq(|\kk|+\sq)}}\\
\df & = \frac{p\sqrt{1+\frac {|\kk|}{\sq}}}{(1-p)\e}.
\end{align}
Despite having multiple observations, it is a valid approach to apply \eqref{diffmse} to this setting, as the reconstruction function only depends on the scalar argument $\overline y_m$. Thus, we have closed form expressions for the reconstruction function \eqref{recfun} and its accuracy in terms of MSE \eqref{msefun}.

Note that for $p=0$, we have
\[
\lim\limits_{p\to 0}\mse(\myk) = \frac{\sigma^2}{|\kk|+\sigma^2}
\]
which does not depend on the observation $\myk$.
Thus, oversampled adaptive sensing (OAS) can achieve a gain only for non-Gaussian signals ($p>0$).

Figure~\ref{figmsefun} 
\begin{figure}
\centerline{\epsfig{file=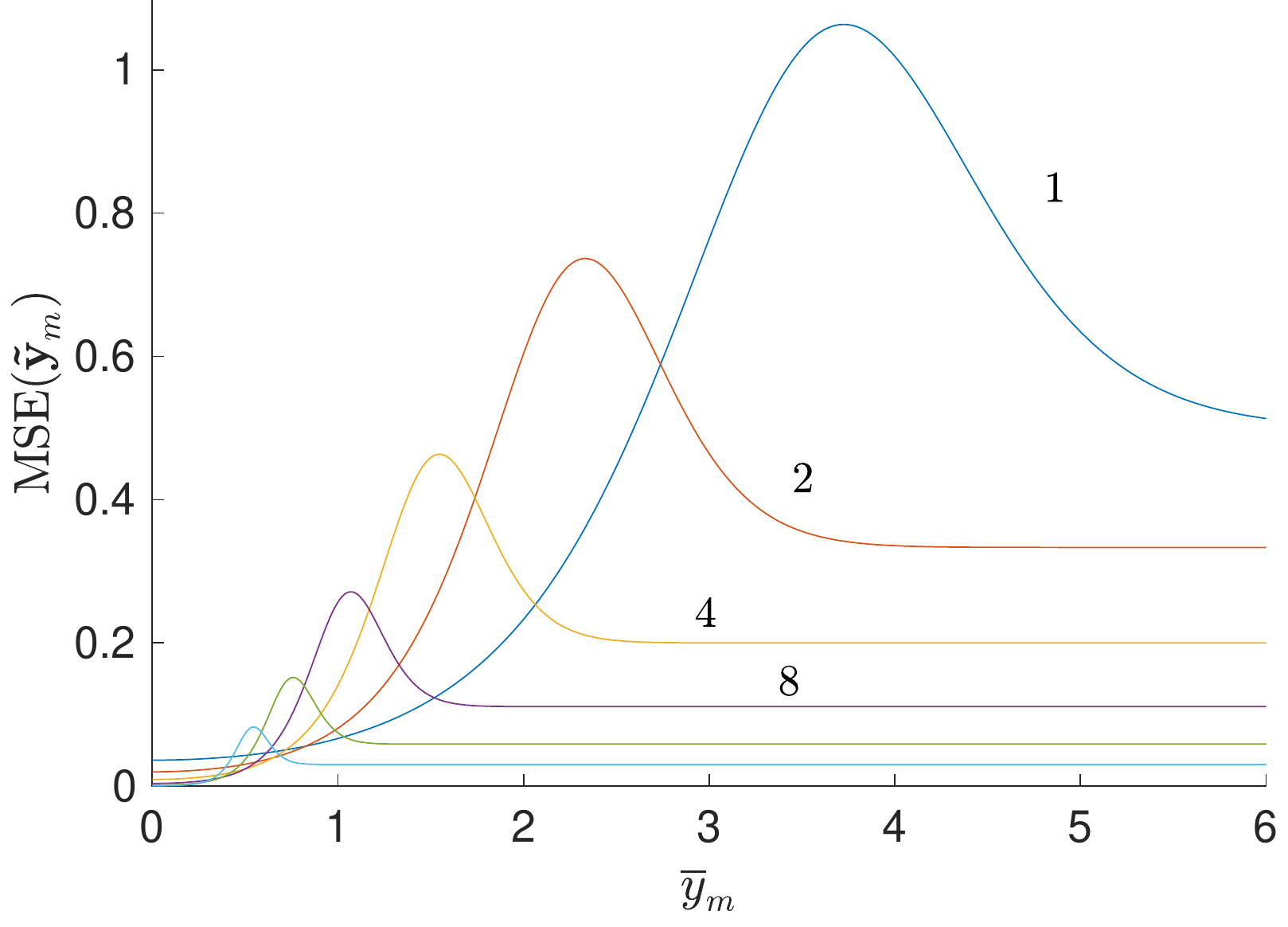,width=.75\columnwidth}}
\caption{\label{figmsefun}
Posterior MSE for sparse Gaussian source with $p=0.9$ and $\sigma^2=1$ for 1, 2, 4, 8, 16, and 32 measurements shown by the highest to lowest curve, resp.}
\end{figure}
illustrates the posterior MSE. 
Note that there exist average values of $\overline y_m$ such that additional observations increase the MSE. The data is a product of a Gaussian and a Bernoulli variable. In a certain range of $\overline y_m$, the error is dominated by wrong binary decisions, i.e.\ taking a zero for a Gaussian or a Gaussian for a zero. The position of that range depends on the strength of the noise and shifts towards zero for increasing number of observations. Thus, it can happen that additional observations shift the position of that range towards the position of the current average observation.  

%%%%%%%%%%%%%%

\subsection{Binary data}
Consider now the case that $\xnk$ follows a binary distribution.
With probability $p$, it is $+1$ and with probability $1-p$ it is $-1$ for all $n_m$.

Specializing \eqref{bayes1} to binary data, we get
\begin{align}
{\rm p}(\xnk|\myk) &= \frac{{\rm e}^{-\sum\limits_{i\in\kk}(\tilde y_i-\xnk)^2/(2\sigma^2)}{\rm p}(\xnk)}{(1-p) {\rm e}^{-\sum\limits_i\frac{(\tilde y_i+1)^2}{2\sigma^2}}  + p e^{-\sum\limits_i\frac{(\tilde y_i-1)^2}{2\sigma^2}}}\\
&= \frac{{\rm e}^{\sum\limits_{i\in\kk} \tilde y_i \xnk/\sigma^2}{\rm p}(\xnk)}{(1-p) {\rm e}^{-\sum\limits_i \tilde y_i/\sigma^2}  + p {\rm e}^{\sum\limits_i\tilde y_i/\sigma^2}}
\label{eq22b}
\end{align}
From \eqref{genrec} and \eqref{diffmse}, we find 
%The optimum function $r(\myk)$ is found by setting the derivative of the MSE to zero:
%\begin{align}
%0&=\frac{\partial \mse(\myk)}{\partial r(\myk)}\\
%&= p \left[ r(\myk)-1\right]{\rm e}^{\sum\limits_{i\in\kk} \tilde y_i /\sigma^2}
%+(1-p) \left[r(\myk)+1\right]{\rm e}^{-\sum\limits_{i\in\kk} \tilde y_i /\sigma^2}
%\end{align}
%This yields
\begin{equation}
\label{funcbb}
r(\myk) =  \frac{
p {\rm e}^{\sum\limits_{i\in\kk} \tilde y_i /\sigma^2} - (1-p) {\rm e}^{-\sum\limits_{i\in\kk} \tilde y_i /\sigma^2} 
}{
p {\rm e}^{\sum\limits_{i\in\kk} \tilde y_i /\sigma^2} + (1-p) {\rm e}^{-\sum\limits_{i\in\kk} \tilde y_i /\sigma^2} 
 }
\end{equation}
and
\begin{align}
\mse(\myk)  & = \frac{
4p(1-p)
}
{
\left[(1-p) {\rm e}^{-\sum\limits_{i\in\kk} \tilde y_i /\sigma^2} +p {\rm e}^{\sum\limits_{i\in\kk} \tilde y_i /\sigma^2}\right]^2
 },
\end{align}
respectively, since the reconstruction function only depends on the sum of the observations.

For equiprobable signals, we get
\be
\lim\limits_{p\to\frac12}\mse(\myk) ={
\cosh^{-2}\left(\sum\limits_{i\in\kk} \tilde y_i /\sigma^2\right)
 }.
\ee
The MSE depends on the observation $\myk$ even for $p=\frac12$. 
Note that the signal is sparse even for $p=\frac12$, as the probability for all data components except for $\pm1$ is zero. Thus, the adaptation of the sensing time is beneficial even for standard binary phase shift keying. In fact, it is well known that feedback  improves the bit error rate, but not the channel capacity. In case of channel capacity, the law of large numbers ensures that all receive symbols have the same reliability. 

Figure~\ref{figbinfun} 
\begin{figure}
\centerline{\epsfig{file=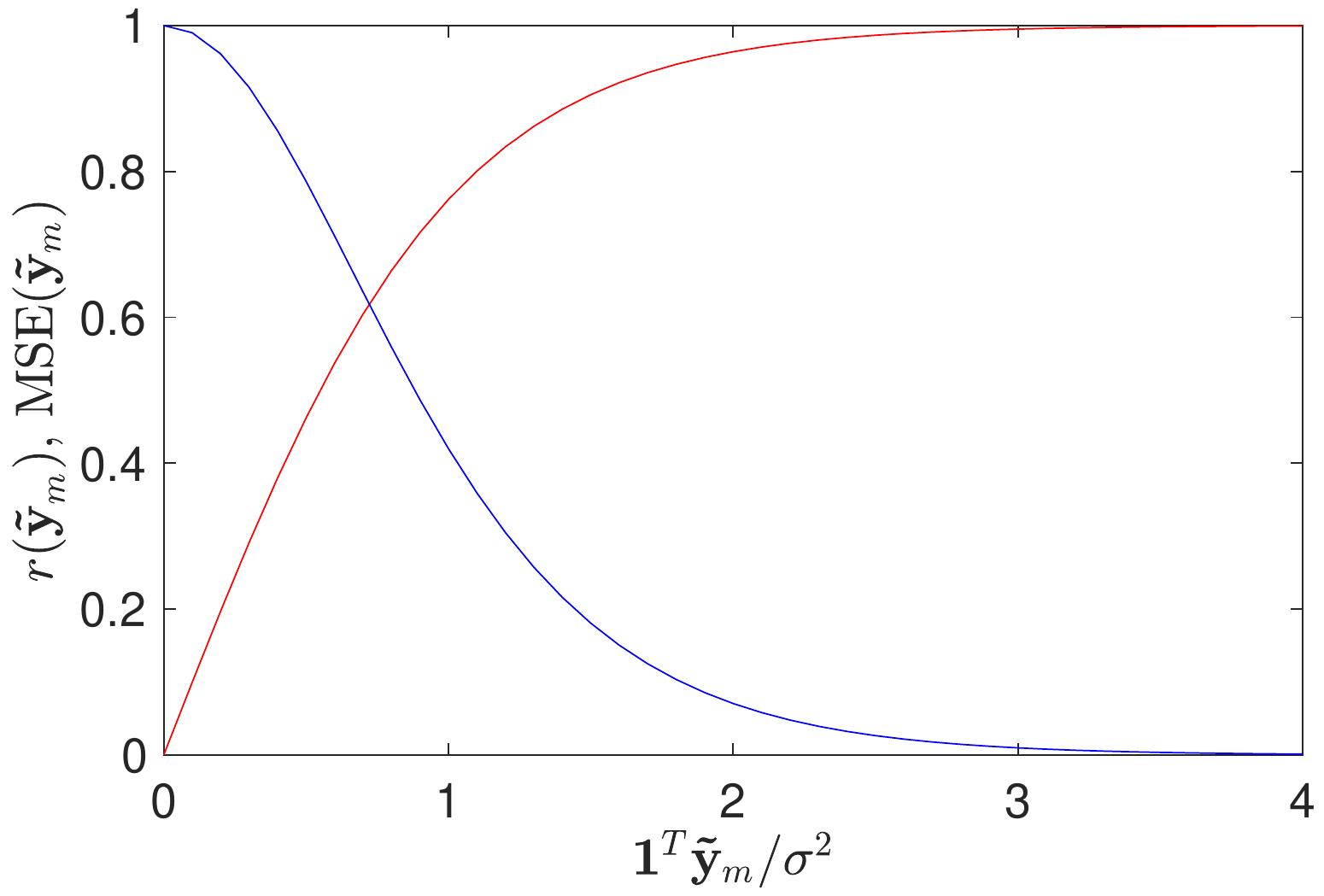,width=.75\columnwidth}}
\caption{\label{figbinfun}
Reconstruction function and posterior MSE for binary source with $p=\frac12$ shown by the red and blue curve, resp.}
\end{figure}
illustrates the reconstruction function and the posterior MSE. Only the positive parts are shown as the reconstruction function and the MSE are odd and even functions, resp. Note that only the sum of the elements, i.e.\ $\matr 1^{\rm T}\myk$ matters.

%%%%%%%%%%%
\subsection{Worst Component Adaptation}
One can use the MSE of the current observation $\myk$ to decide which data to measure next, i.e.\ how to choose $n_{m+1}$.
Here the adaptive nature of OAS comes into play.
At the beginning, each data component of $\matr x$ will be measured once.
Thus, the choices $n_m=m$ and $\kk = \{m\}$ are natural for all $m\le N$. However, the $N+1^{\rm st}$ measurement is taken from the data component 
\be
n_{N+1} = \argmax\limits_{m\le N} \mse(\myk)
\ee
for which the posterior $\mse(\myk)$ is largest. Thus, we get ${\cal M}_{N+1} = \{n_{N+1}, N+1 \}$.
In general, we have the recursion that for any given $m$, the next measurement is taken from the data component 
\be
 n_{m+1}= n_{\ell_m} 
\ee
with\footnote{The condition ${\cal M}_{m^\prime} \not\subset {\cal M}_{i}\forall i$ excludes former measurements, which have already been improved, from the search for the currently worst measurement.} 
\be \ell_m = \argmax\limits_{m^\prime \le m: {\cal M}_{m^\prime} \not\subset {\cal M}_{i}\forall i} \mse(\myk).
\ee
Thus, the adaptive measurements follow the recursion
\be
{\cal M}_{m+1} = {\cal M}_{\ell_m} \cup \{m+1\}.
\ee
Note that the next measurement is always chosen from that component  of the data with the largest posterior MSE. This holds irrespective of how often this component has been measured already. For some components of the data, the realizations of the noise samples may have turned out more hostile than for other data components. These data components will be measured more frequently.

%%%%%%%%%%%%%%%%%%
\subsection{Asymptotic Adaptation}
\label{aa}

From a practical point of view, it is very inconvenient to switch from one data component to another at frequencies as high as $1/T=M/\Tm$.
Somehow counterintuitively, this drawback can be overcome by increasing the oversampling factor $M$.

The smaller the time increment $T$, the smaller the difference in MSE due to a new measurement. Thus, we have
\be
\lim\limits_{M\to\infty} |\mse(\matt y_{m+1})-\mse(\myk)| = 0. 
\ee
Note that the MSE does not converge to zero for infinite oversampling factor, as the noise variance $\sigma^2=MN_0\Tm$ scales with the oversampling factor.
For very large oversampling factor, i.e.\ very fine granularity of measurement time, all data components will be measured with very similar accuracy, i.e.\ they have almost the same posterior MSE.

In order to achieve a certain target mean squared error, we need not continuously check which data component to measure next. We simply measure the current one just as long until the required MSE has been reached.
Then, we continue with the next component.
The target MSE determines the average measurement time.

Asymptotic adaptation leads to measurement schedules were the same component is measured repeatedly for a longer period of time without any adaptation taking place. The average duration of that time is $\Tm=MT$. This is in contrast to worst component adaptation where the measured component typically changes every measurement period of duration $T$.

Asymptotic adaptation allows for $K$ measurements to be fully taken in parallel. 
At the beginning, the first $K$ components of the data vector $\matr x$ are sensed until the MSE of one of these components reaches the desired threshold.
Then, one sensor has become free. This sensor is then used to measure the $K+1^{\rm st}$ component of the data vector.
Whenever, one sensor becomes free, because it has reached the desired MSE, it is used to measure the next still unmeasured component of the data vector.
If no unmeasured data component is left over, the free sensor is used to measure that data component which has the largest current MSE.
In this termination phase of the measurements, more and more sensors measure fewer and fewer data components simultaneously. 
Thus, the row weight of the measurement matrix increases from 1 to $K$. How to deal with non-unit rows weights is addressed in the sequel.

\section{Multiple Rows Weight Measurements}

Adaptive sensing based on oversampling can be combined with classical ideas of compressive sensing. This approach is detailed in this section.

Consider now a sequence of measurement matrices $\matr A_m\in\R^{K\times N}$ such that at measurement time $m$, we measure
\be
\matr y_m = \matr A_m \matr x + \matr z_m.
\ee
We collect all previous observations $\matr y_1,\dots, \matr y_M$, and noise realizations $\matr z_m$ into matrices to obtain
\be
\matr Y_m = [\matr A_1\matr x,\dots,\matr A_m\matr x] +\matr Z.
\ee
We introduce individual reconstruction functions $r_n(\matr Y_m)$ for all $N$ components of the source vector $\matr x$ and individual conditional average distortions
\be
D_n(\matr Y_m) = \expect\limits_{\matr x|\matr Y_m} d[x_n,r_n(\matr Y_m)].
\ee
The optimal reconstruction functions can be found by Bayesian estimation. In practice, one would often use suboptimal ones based on LASSO regression \cite{tibshirani:96,chen:01} or approximate message passing \cite{donoho:09,montanari:12,ma:17}.

In addition to the reconstruction functions, the conditional average distortions need to be evaluated. This can also be performed by means of, e.g., approximate message passing.
In case of Bayesian reconstruction and a large number of source components, i.e.\ $N\gg1$, the conditional average distortion may be approximated by means of \eqref{diffmse}, if the vector-estimation problem asymptotically decouples into scalar  estimation problems with equivalent Gaussian noise channels. Detailed conditions for such a behavior can be found in \cite{bereyhi:16}.
 
%%%
A sensible adaption criterion is based on the asymptotic adaptation rule outlined in Section~\ref{aa}. If the conditional average distortion $D_n(\matr Y_m)$ of some component $x_n$ falls below a given target distortion, this component is discontinued to be sensed.
Thus, the $n$-th columns of future sensing matrices are nulled. This improves the quality of future measurements due to a reduced instantaneous compression ratio.
 
%%%%%%%%%%%%%%%%%%%%
\section{Implementation Aspects}

The decision which component to measure next is based upon the instantaneous measurement fidelity expressed in terms of the instantaneous distortion measure.
Even for Gaussian sparse sources with mean-square distortion, calculation of the instantaneous MSE requires the evaluation of \eqref{msefun}.
For other source statistics and/or other distortion measures, a closed form expression for the instantaneous distortion might even not exist.
For the feasibility of OAS in practice, an efficient implementation of the measurement scheduler is very important.

The instantaneous MSE for a sparse Gaussian source is depicted in Fig.~\ref{figmsefun}. 
Irrespective of the number of measurements, all curves show the same behavior. 
The functions monotonically increase until they reach their maxima. Then, they decrease monotonically.
For sources with other statistics, the functions might be more complicated. Several local maxima might occur. 

Irrespective of the precise shape of the instantaneous fidelity functions, there is no need to evaluate them more than a single time, if asymptotic adaptation is applied.
Any given target MSE is achieved by a certain subset of the x-axis in Fig.~\ref{figimplem}.
\begin{figure}
\centerline{\epsfig{file=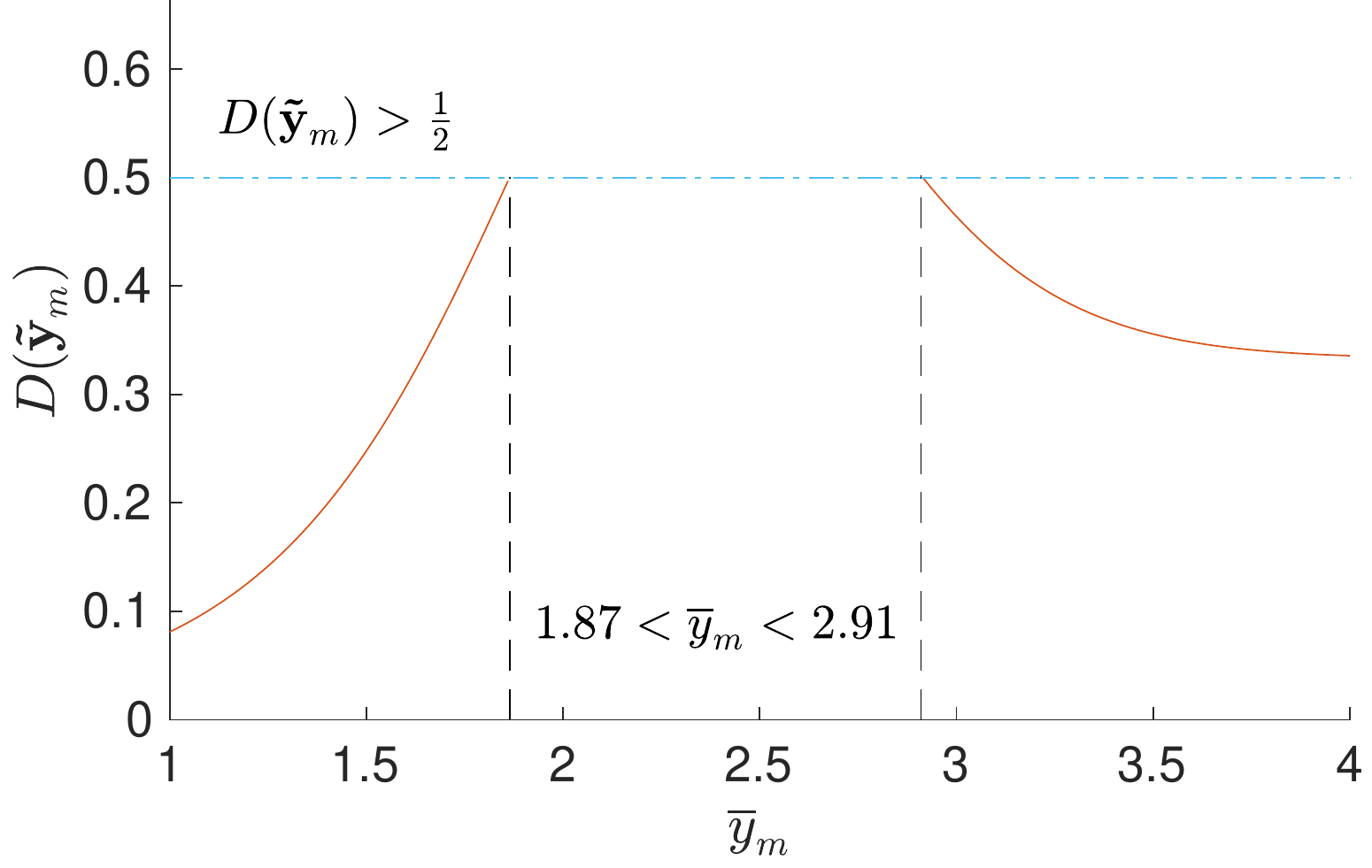,width=.75\columnwidth}}
\caption{\label{figimplem}
Transformation of a distortion-based stopping criterion to an observation based one.}
\end{figure}
For any number of measurements, the respective subsets can be calculated (or simulated, if analytical calculations do not work out for the source statistics) before the measurements have started.
If $\matr 1^{\rm T}\myk$ falls within the respective subset, measurements are terminated.
For Gaussian sparse sources and mean-square distortion, this procedure results in only two comparisons of $|\overline y_m|$: one against a lower and one against and an upper threshold.

The threshold comparison is a very simple operation. It can even be performed by the analog part of the ADC. In that case, time need not be quantized, i.e.\ $M\to\infty$, and the asymptotic adaptation rule is equivalent to worst component adaptation.

For whatever reasons, the threshold comparison might be implemented in discrete-time hardware.
The power consumption of ADCs is well-known to scale linearly with sampling rate. However, this scaling law does not apply to OAS.
ADCs charge one or more capacitors once per sampling period. Before the next sample is taken, the capacitors are discharged, i.e.\ the stored energy is converted into heat.
In case of OAS, the capacitors need not be discharged, but the charge may accumulate as the reconstruction function only depends on the sum of all samples, see e.g. \eqref{defaverage} for sparse Gaussian sources. Thus, for OAS, the power consumption does not scale linearly with the oversampling factor $M$, if it scales at all.

%%%%%%%%%%%%%%%%%
\section{Comparison to Nonadaptive CS}
In this section, we compare OAS against the following two states of the arts. 
\begin{enumerate}
\item
Orthogonal sensing with reduced sampling time. In order to get compression factor $N/K$, the measurement time per sample is reduced from $\Tm$ to $\Tm K/N$.
\item
CS with a random sensing matrix of size $K\times N$ whose entries are either independent and identically distributed (iid) or follow a Haar distribution.
\end{enumerate}
Numerical results are given in Fig.~\ref{figali}.
\begin{figure}
\centerline{\epsfig{file=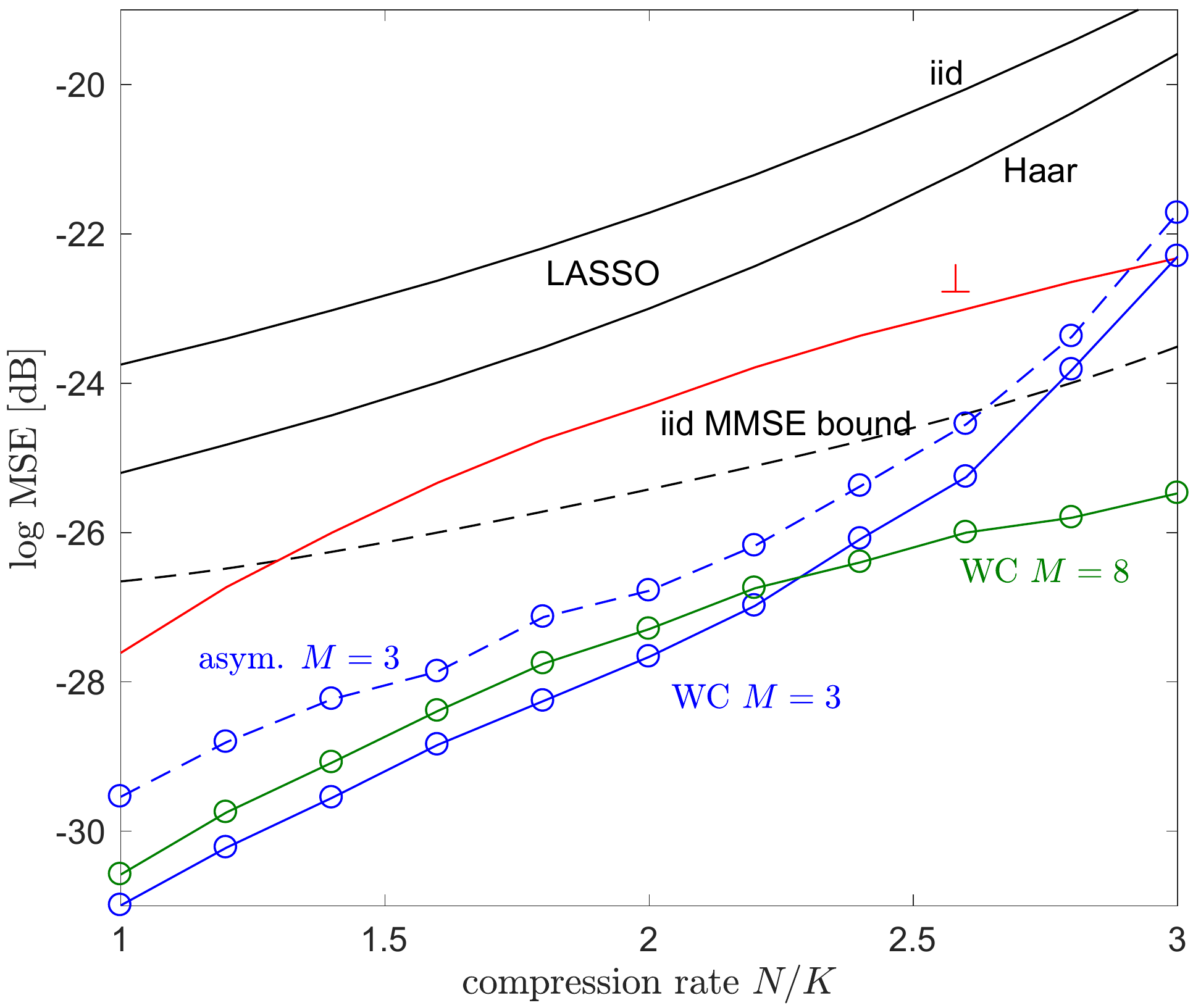,width=.925\columnwidth}}
\caption{\label{figali}
MSE vs.\ compression ratio for a sparse Gaussian source with $p=0.9$ and $\log E_{\rm s}/N_0 = 10$ dB.
OAS is simulated for the worst-component (WC) adaptation and the asymptotic adaptation rule with unit row weight measurements and $N=100$ averaged over $10^4$ realizations. The results for classical CS are analytic large-system results according to \cite{bereyhi:18}.}
\end{figure}
Not surprisingly, LASSO falls somewhat behind the minimum MSE bound and iid sensing matrices perform somewhat worse than Haar distributed ones.
Orthogonal sensing with reduced sampling time is inferior to classical CS approaches over a wide range of compression ratios.
OAS shows superior performance to all other approaches as long as the oversampling factor exceeds the compression rate. The asymptotic adaptation rule falls slightly behind worst component adaptation. It is unclear so far, why the performance does not monotonically increase with the oversampling factor. One reasons could be that none of the discussed adaptation rules is the optimum one.
To clarify this issue, further investigations are necessary.
Simulations for multiple-row weight measurements are ongoing and were not finished when this manuscript was typeset.
%%%%%%%%%%%%%%%%%%%%
\section{Conclusions}
OAS is a viable alternative to CS based on random linear measurements. 
Adaptation of the sensing matrix to preliminary measurement results offers significant potential for improvement of the fidelity of reconstruction.
For adaptive sensing, the RIP is not required, as orthogonal measurement were shown to perform excellently. OAS is well suited for hardware implementation.

%%%%%%%%%%%%%%%%%
%\section{Bibliography}
%
\bibliographystyle{unsrt}
\bibliography{lit}

\end{document}